# Passive Power Filters

*R. Künzi*
Paul Scherrer Institute, Villigen, Switzerland

**Abstract**
Power converters require passive low-pass filters which are capable of reducing voltage ripples effectively. In contrast to signal filters, the components of power filters must carry large currents or withstand large voltages, respectively. In this paper, three different suitable filter structures for d.c./d.c. power converters with inductive load are introduced. The formulas needed to calculate the filter components are derived step by step and practical examples are given. The behaviour of the three discussed filters is compared by means of the examples. Practical aspects for the realization of power filters are also discussed.

## 1 Introduction

Switched mode d.c./d.c. power converters very often have the structure shown in Fig. 1.

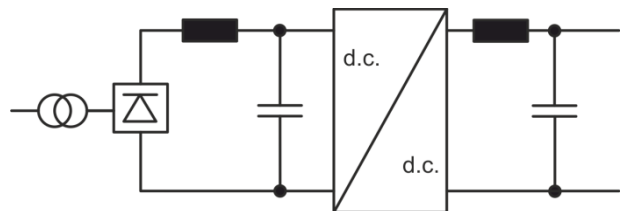

**Fig. 1:** General structure of a switched mode converter

The input transformer and rectifier form a non-controlled d.c.-link voltage with a rather large voltage ripple. An input low-pass filter is needed to reduce this voltage ripple. This filter deals with voltage ripples of typically six times the mains frequency and higher-order harmonics of that. Therefore the components must be designed for frequencies up to a few kilohertz. The capacitor also acts as a voltage source for the d.c./d.c. converter and therefore has to carry large currents at high frequencies. Such input filters require rather large components, the requirements regarding high-frequency behaviour are moderate.

Usually, an output filter is also required to filter the output voltage ripple. An inductive load acts itself as a very effective filter for the current. However, without the output filter the cabling between the converter and the load would carry large a.c. voltages with large voltage slopes. That can cause large electro-magnetic interferences and the cables would need to be shielded. The fundamental frequency of the voltage ripple is equal to the switching frequency of the converter or a multiple of that. High-stability converters for accelerator applications must have a high closed-loop bandwidth in order to react to errors quickly enough. The output filter limits the closed-loop bandwidth; its cut-off frequency must therefore be as high as possible. On the other hand, the cut-off frequency must be well below the switching frequency to reduce the ripple voltage effectively. This leads to converters with rather high switching frequencies of several tens of kilohertz combined with high-order output filters. The filter components must therefore be designed for frequencies up to a few hundred kilohertz.

Various structures for passive low-pass filters are listed and evaluated in Table 1 for their suitability for power converters.

**Table 1:** Suitable filter structures

| Circuit | Description |
|---|---|
| *RC low-pass* | For signal filters simple *RC* circuits are commonly used. They offer an attenuation of only 20 dB/decade. The full current flows through the resistor, which causes high losses.<br><br>Not suitable! |
| *LC low-pass* | With an *LC* structure we get 40 dB/decade, but there is a large resonance!<br><br>Not suitable! |
| *LC with series R* | Series damping in order to overcome the resonance problem: the full current flows through the resistor. If the parasitic resistance of the inductor is large enough, this might be ok.<br><br>Usually not suitable! |
| *LC with parallel R* | Parallel damping in order to overcome the resonance problem: the full voltage is across the resistor, which causes high losses.<br><br>Not suitable! |
| *LC with parallel RC* | Parallel *RC* damping in order to overcome the resonance problem: the resonance can be damped effectively and the losses are reasonable.<br><br>Suitable; see Section 2. |
| *Two LC stages* | Two *LC* stages offer an attenuation of 80 dB/decade, but there is again the resonance problem.<br><br>Not suitable! |
| *Two LC with parallel RC in first stage* | Parallel *RC* damping in the first stage in order to overcome the resonance problem: losses are moderate.<br><br>Suitable, but not optimal; see Section 3.1. |
| *Two LC with parallel RC in second stage* | Parallel *RC* damping in the second stage in order to overcome the resonance problem: losses are low.<br><br>Suitable; see Section 3.2. |

## 2 Design of a second-order low-pass filter

In this section a low-pass filter according to Fig. 2 is outlined.

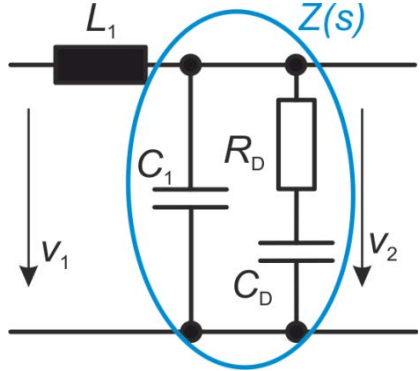

**Fig. 2:** Second-order low-pass filter

The complex impedance $Z(s)$ is

$$Z(s) = \cfrac{1}{C_1 s + \cfrac{1}{R_\mathrm{D} + \cfrac{1}{C_\mathrm{D} s}}} = \cfrac{1}{C_1 s + \cfrac{C_\mathrm{D} s}{R_\mathrm{D} C_\mathrm{D} s + 1}} = \cfrac{R_\mathrm{D} C_\mathrm{D} s + 1}{C_1 R_\mathrm{D} C_\mathrm{D} s^2 + (C_1 + C_\mathrm{D}) s}. \tag{1}$$

The transfer function of the entire filter is then

$$G(s) = \frac{v_2(s)}{v_1(s)} = \frac{Z(s)}{L_1 s + Z(s)} = \cfrac{\cfrac{R_\mathrm{D} C_\mathrm{D} s + 1}{C_1 R_\mathrm{D} C_\mathrm{D} s^2 + (C_1 + C_\mathrm{D}) s}}{L_1 s + \cfrac{R_\mathrm{D} C_\mathrm{D} s + 1}{C_1 R_\mathrm{D} C_\mathrm{D} s^2 + (C_1 + C_\mathrm{D}) s}}$$

$$= \frac{R_\mathrm{D} C_\mathrm{D} s + 1}{L_1 C_1 R_\mathrm{D} C_\mathrm{D}\, s^3 + L_1 (C_1 + C_\mathrm{D}) s^2 + R_\mathrm{D} C_\mathrm{D} s + 1}. \tag{2}$$

Note that the *s* terms in the numerator and in the denominator are equal. Therefore we can write Eq. (2) as

$$G(s) = \frac{k_1 s + 1}{k_3 s^3 + k_2 s^2 + k_1 s + 1} \tag{3a}$$

$$\text{with}\quad k_1 = R_\mathrm{D} C_\mathrm{D}, \tag{3b}$$

$$k_2 = L_1 (C_1 + C_\mathrm{D}), \tag{3c}$$

$$k_3 = L_1 C_1 R_\mathrm{D} C_\mathrm{D}. \tag{3d}$$

$G(s)$ can be expressed as the combination of a third-order PT (time-delay element) in the denominator and a first-order PD (proportional plus derivative element) in the numerator. Let us have a closer look at the third-order PT. This can be split into one second-order and one first-order PT, both connected in series. It can be expressed in its normalized form as

$$G_\mathrm{PT}(s) = \frac{1}{\left(1 + a_1 \dfrac{s}{\omega_0}\right) \cdot \left(1 + a_2 \dfrac{s}{\omega_0} + b_2 \dfrac{s^2}{\omega_0^2}\right)}. \tag{4}$$

The transfer function of such a combination of first- and second-order PTs can be optimized according to different methods [1], which results in particular values for $a_i$ and $b_i$. The commonly used optimization methods with their corresponding coefficients are given in Table 2.

Table 2: Coefficients for a third-order PT for different optimization methods

| Method | $a_1$ | $a_2$ | $b_2$ |
| --- | --- | --- | --- |
| Butterworth | 1.0000 | 1.0000 | 1.0000 |
| Bessel | 0.7560 | 0.9996 | 0.4772 |
| Critical damping | 0.5098 | 1.0197 | 0.2599 |

By expanding Eq. (4) we get

$$G_{\mathrm{PT}}(s) = \frac{1}{1 + \frac{a_2}{\omega_0}s + \frac{b_2}{\omega_0^2}s^2 + \frac{a_1}{\omega_0}s + \frac{a_1 a_2}{\omega_0^2}s^2 + \frac{a_1 b_2}{\omega_0^3}s^3}$$

$$= \frac{1}{\frac{a_1 b_2}{\omega_0^3}s^3 + \frac{(a_1 a_2 + b_2)}{\omega_0^2}s^2 + \frac{(a_1 + a_2)}{\omega_0}s + 1}. \tag{5}$$

By comparing the coefficients with Eq. (3a), we get

$$k_1 = R_\mathrm{D} C_\mathrm{D} = \frac{a_1 + a_2}{\omega_0}, \tag{6a}$$

$$k_2 = L_1(C_1 + C_\mathrm{D}) = \frac{a_1 a_2 + b_2}{\omega_0^2}, \tag{6b}$$

$$k_3 = L_1 C_1 R_\mathrm{D} C_\mathrm{D} = \frac{a_1 b_2}{\omega_0^3}. \tag{6c}$$

For a given optimization method, the three independent Eqs. (6a)–(6c) contain five unknowns ($L_1$, $C_1$, $R_\mathrm{D}$, $C_\mathrm{D}$, and $\omega_0$). Therefore, we have the choice to select two of them and the remaining three depend on that selection.

Selection of the cut-off angular frequency $\omega_0$: for a given angular frequency $\omega_\mathrm{B}$ well in the blocking area of the filter ($\omega_\mathrm{B} \gg \omega_0$) we can define the desired attenuation $G_\mathrm{B}$. In the blocking area the highest-order terms of both the numerator and the denominator in Eq. (2) dominate; therefore it can be simplified to

$$G_\mathrm{B} = \frac{R_\mathrm{D} C_\mathrm{D} s}{L_1 C_1 R_\mathrm{D} C_\mathrm{D} s^3} = \frac{\frac{a_1 + a_2}{\omega_0}s}{\frac{a_1 b_2}{\omega_0^3}s^3} = \frac{a_1 + a_2}{a_1 b_2} \cdot \frac{\omega_0^2}{s^2} = \frac{a_1 + a_2}{a_1 b_2} \cdot \frac{\omega_0^2}{\omega_\mathrm{B}^2},$$

$$\omega_0 = \omega_\mathrm{B} \cdot \sqrt{\frac{G_\mathrm{B} \cdot a_1 b_2}{a_1 + a_2}}. \tag{7}$$

Selection of $C_1$: if $C_1$ serves also as a commutation capacitor of a converter, it carries large a.c. currents. Therefore its capacitance must often be selected according to the current capability, in order to limit the temperature rise and to prevent early aging.

Selection of $L_1$: the inductance $L_1$ should be optimized for a reasonable ripple current. For cost reasons $L_1$ should be as low as possible, but a too low inductance results in an excessive ripple current. As an example, Fig. 3 shows the inductor ripple current for a buck converter.

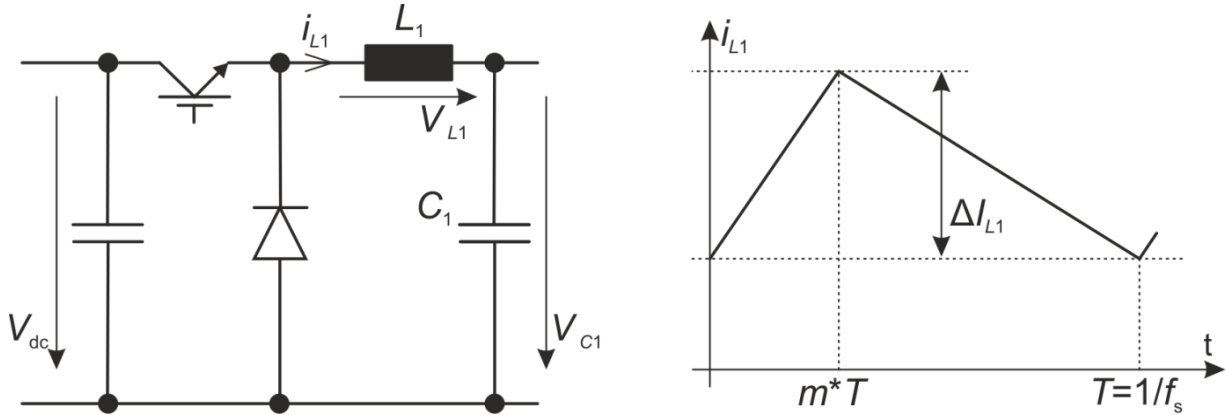

**Fig. 3:** Ripple current in the inductor $L_1$ of a buck converter

If $m$ is the duty cycle of the switch, the d.c. voltage across $C_1$ is $m*V_{dc}$. When the switch is on, the current in $L_1$ increases and the peak–peak ripple current $\Delta I_{L1}$ can be calculated as

$$V_{L1} = L_1 \cdot \frac{di_{L1}}{dt} = V_{dc} - V_{C1} = V_{dc} \cdot (1 - m),$$

$$\Delta I_{L1} = m \cdot T \cdot \frac{di_{L1}}{dt} = m \cdot \frac{1}{f_s} \cdot \frac{V_{dc} \cdot (1-m)}{L_1} = \frac{V_{dc} \cdot (1-m) \cdot m}{f_s \cdot L_1}.$$

The function $(1 - m)*m$ has its maximum of 0.25 at $m = 0.5$. Therefore, $L_1$ can be calculated as:

$$L_1 = \frac{V_{dc} \cdot 0.25}{f_s \cdot \Delta I_{L1}} \qquad (8a)$$

Figure 4 illustrates an alternative approach to determine $L_1$ for sinusoidal ripple currents and voltages.

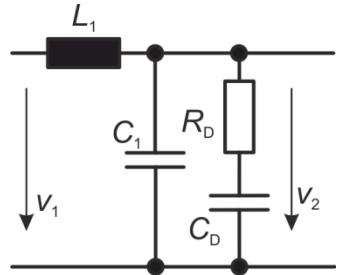

**Fig. 4:** Ripple current in the inductor $L_1$ for sinusoidal ripple currents and voltages

The filter eliminates the voltage ripple of $v_1$ nearly entirely, so the voltage ripple in $v_2$ is much smaller. Therefore, the a.c. ripple of $v_1$ (with frequency $f_1$) is also present across $L_1$ and generates a ripple current in $L_1$:

$$I_{L1\_ripple\_peak\_peak} = \frac{v_{1\_ripple\_peak\_peak}}{2 \cdot \pi \cdot f_1 \cdot L_1},$$

$$L_1 = \frac{v_{1\_ripple\_peak\_peak}}{2 \cdot \pi \cdot f_1 \cdot I_{L1\_ripple\_peak\_peak}}. \qquad (8b)$$

As mentioned before, we have the choice to preselect two of the three parameters $C_1$, $L_1$, and $\omega_0$. By substituting Eq. (6a) into Eq. (6c) we get an equation that can be solved for the remaining parameter:

$$C_1 = \frac{a_1 b_2}{L_1 \omega_0^2 (a_1 + a_2)}, \tag{9a}$$

$$L_1 = \frac{a_1 b_2}{C_1 \omega_0^2 (a_1 + a_2)}, \tag{9b}$$

$$\omega_0 = \sqrt{\frac{a_1 b_2}{L_1 C_1 (a_1 + a_2)}}. \tag{9c}$$

By solving Eq. (6b) for $C_D$ we get

$$C_D = \frac{a_1 a_2 + b_2}{L_1 \omega_0^2} - C_1. \tag{10}$$

By solving Eq. (6a) for $R_D$ we get

$$R_D = \frac{a_1 + a_2}{C_D \omega_0}. \tag{11}$$

Example 1: Design a second-order filter, which will be placed between a diode rectifier and a buck converter according to Fig. 5 and will reduce the 300 Hz ripple voltage from the rectifier bridge. The d.c.-link voltage is 200 V and the ripple current in $L_1$ must not exceed 50 A peak to peak. The design of the buck converter has shown that $C_1$ needs to be 22 mF to get a reasonable capacitor current. Make the design for all three given optimization methods and compare the results.

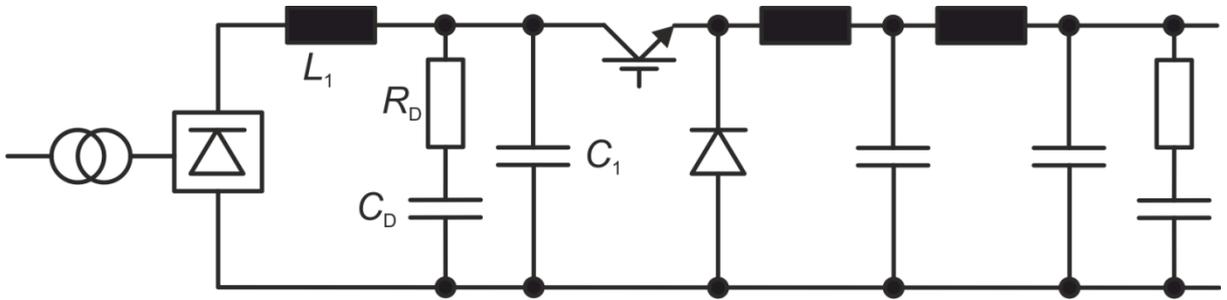

**Fig. 5:** Input filter for a buck converter

The diode rectifier produces a 300 Hz voltage ripple of approximately 13% of 200 V, i.e., 26 V peak to peak. To keep it simple, we consider this ripple to be sinusoidal. That means we can determine $L_1$ according to Eq. (8b) as follows:

$$L_1 = \frac{v_{1\_\text{ripple\_peak\_peak}}}{2 \cdot \pi \cdot f_1 \cdot I_{L1\_\text{ripple\_peak\_peak}}} = \frac{26 \text{ V peak to peak}}{2 \cdot \pi \cdot 300 \text{ s}^{-1} \cdot 50 \text{ A peak to peak}} = 276 \ \mu\text{H}.$$

Select $L_1 = 300 \ \mu\text{H}$ and $C_1 = 22$ mF and calculate the remaining filter components by using Eqs. (9c), (10), and (11). The results are listed in Table 3.

**Table 3:** Results for Example 1

| Parameter | Butterworth | Bessel | Critical damping |
|---|---|---|---|
| $\omega_0$ | 275 s$^{-1}$ | 177 s$^{-1}$ | 115 s$^{-1}$ |
| $f_0$ | 44 Hz | 28 Hz | 18 Hz |
| $L_1$ | 300 µH | 300 µH | 300 µH |
| $C_1$ | 22 mF | 22 mF | 22 mF |
| $C_D$ | 66 mF | 110 mF | 176 mF |
| $R_D$ | 0.11 Ω | 0.09 Ω | 0.08 Ω |

Figure 6 shows the Bode plots of the three filter designs according to Example 1. The three designs differ only around the cut-off frequency. In the Butterworth optimization $C_D$ is minimal but with the drawback of a high resonance gain of 4.5 dB. With critical damping, this resonance gain is reduced to 2.3 dB with the drawback of a large $C_D$. The Bessel optimization is between the two and could be a good compromise.

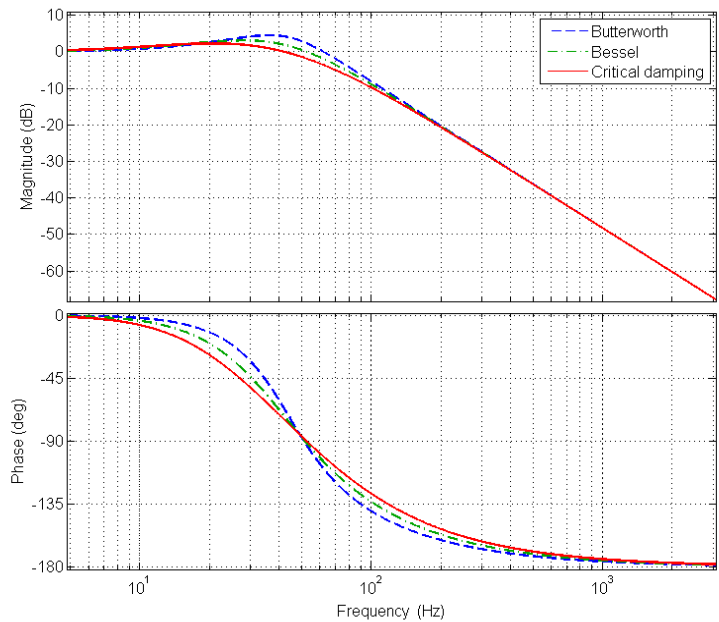

**Fig. 6:** Bode plots for the filter designs according to Example 1

Example 2: Design a second-order filter for a buck converter with a d.c.-link voltage of 120 V, a switching frequency of 20 kHz, and a maximum output current of 500 A; refer to Fig. 7. The ripple current in $L_1$ should not exceed 50 A peak to peak. The filter should have an attenuation of 0.004 at the switching frequency. Design the filter for all three optimization methods and compare the results.

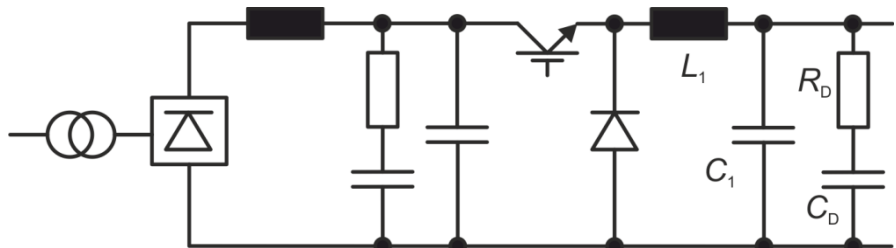

**Fig. 7:** Output filter for a buck converter

Select $L_1$ in order to meet the ripple requirement by using Eq. (8a):

$$L_1 = \frac{V_{\text{dc}} \cdot 0.25}{f_s \cdot \Delta I_{L1}} = \frac{120 \text{ V} \cdot 0.25}{20 \text{ kHz} \cdot 50 \text{ A}} = 30 \text{ } \mu\text{H}.$$

Select $\omega_0$ in order to meet the attenuation requirement by using Eq. (7):

$$\omega_0 = \omega_B \cdot \sqrt{\frac{G_B \cdot a_1 b_2}{a_1 + a_2}} = 2 \cdot \pi \cdot 20 \text{ kHz} \cdot \sqrt{\frac{0.004 \cdot a_1 b_2}{a_1 + a_2}}.$$

Calculate the remaining filter components by using Eqs. (9a), (10), and (11). The results are listed in Table 4.

**Table 4:** Results for Example 2

| Parameter | Butterworth | Bessel | Critical damping |
|---|---|---|---|
| $\omega_0$ | 5620 s$^{-1}$ | 3600 s$^{-1}$ | 2340 s$^{-1}$ |
| $f_0$ | 890 Hz | 570 Hz | 370 Hz |
| $L_1$ | 30 $\mu$H | 30 $\mu$H | 30 $\mu$H |
| $C_1$ | 528 $\mu$F | 528 $\mu$F | 528 $\mu$F |
| $C_D$ | 1580 $\mu$F | 2640 $\mu$F | 4220 $\mu$F |
| $R_D$ | 0.22 $\Omega$ | 0.18 $\Omega$ | 0.15 $\Omega$ |

Figure 8 shows the Bode plots of the three filter designs according to Example 2. The Bode plots are similar to the ones for Example 1 except for the frequency scaling. The attenuation at 20 kHz is −48 dB, which corresponds to a factor of 0.004, as required. The same trade-off between small capacitor values and low resonance amplitude applies; refer to Example 1.

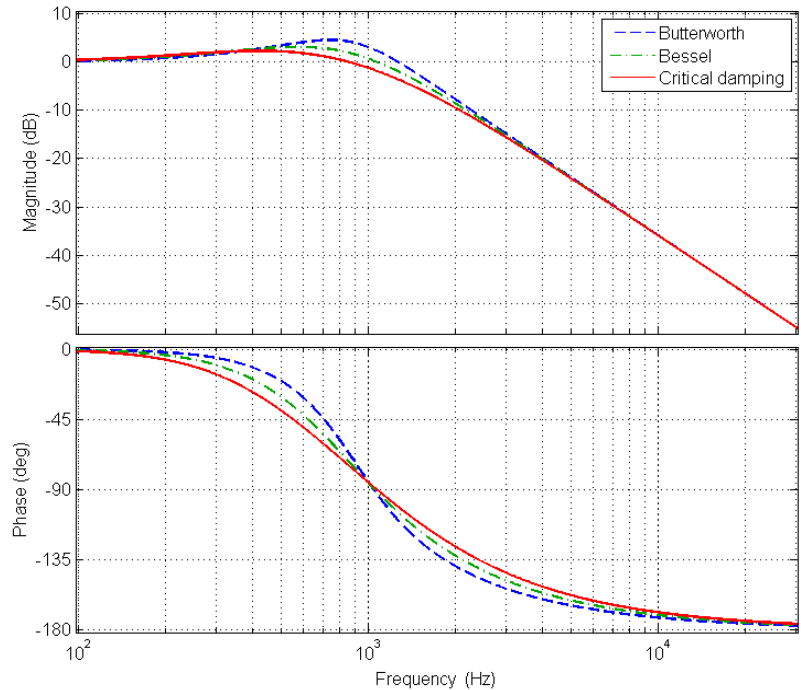

**Fig. 8:** Bode plots for the filter designs according to Example 2

## 3 Design of a fourth-order low-pass filter

For the fourth-order low-pass filter there are two alternatives. One of them has the *RC*-damping circuit in the first, the other one in the second *LC* stage. Both alternatives are outlined in detail in Sections 3.1 and 3.2, respectively, and are compared with each other in Section 4.

### 3.1 Fourth-order low-pass filter with damping circuit in the first *LC* stage

In this section a low-pass filter according to Fig. 9 is outlined.

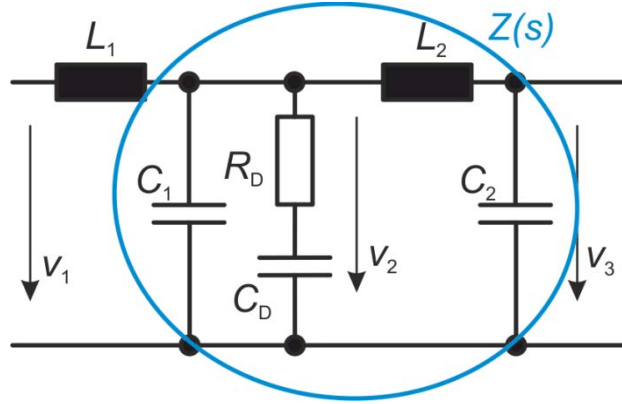

**Fig. 9:** Fourth-order low-pass filter with *RC* damping in the first *LC* stage

The complex impedance $Z(s)$ is

$$Z(s) = \cfrac{1}{C_1 s + \cfrac{1}{R_D + \cfrac{1}{C_D s}} + \cfrac{1}{L_2 s + \cfrac{1}{C_2 s}}} = \cfrac{1}{C_1 s + \cfrac{C_D s}{R_D C_D s + 1} + \cfrac{C_2 s}{L_2 C_2 s^2 + 1}}$$

$$= \frac{(L_2 C_2 s^2 + 1) \cdot (R_D C_D s + 1)}{C_1 s \cdot (L_2 C_2 s^2 + 1) \cdot (R_D C_D s + 1) + C_D s \cdot (L_2 C_2 s^2 + 1) + C_2 s \cdot (R_D C_D s + 1)}$$

$$= \frac{(L_2 C_2 s^2 + 1) \cdot (R_D C_D s + 1)}{L_2 C_1 C_2 R_D C_D s^4 + L_2 C_1 C_2 s^3 + C_1 R_D C_D s^2 + C_1 s + L_2 C_2 C_D s^3 + C_D s + C_2 R_D C_D s^2 + C_2 s};$$

$$Z(s) = \frac{(L_2 C_2 s^2 + 1) \cdot (R_D C_D s + 1)}{L_2 C_1 C_2 R_D C_D s^4 + L_2 C_2 (C_1 + C_D) s^3 + R_D C_D (C_1 + C_2) s^2 + (C_1 + C_2 + C_D) s}. \tag{12}$$

The transfer function can be split into two partial transfer functions $G_1(s)$ and $G_2(s)$:

$$G(s) = G_1(s) \cdot G_2(s) = \frac{v_2(s)}{v_1(s)} \cdot \frac{v_3(s)}{v_2(s)} = \frac{v_3(s)}{v_1(s)};$$

$$G_1(s) = \frac{v_2(s)}{v_1(s)} = \frac{Z(s)}{L_1 s + Z(s)}$$

$$= \cfrac{\cfrac{(L_2 C_2 s^2 + 1) \cdot (R_D C_D s + 1)}{L_2 C_1 C_2 R_D C_D s^4 + L_2 C_2 (C_1 + C_D) s^3 + R_D C_D (C_1 + C_2) s^2 + (C_1 + C_2 + C_D) s}}{L_1 s + \cfrac{(L_2 C_2 s^2 + 1) \cdot (R_D C_D s + 1)}{L_2 C_1 C_2 R_D C_D s^4 + L_2 C_2 (C_1 + C_D) s^3 + R_D C_D (C_1 + C_2) s^2 + (C_1 + C_2 + C_D) s}}$$

$$= \frac{(L_2 C_2 s^2 + 1) \cdot (R_D C_D s + 1)}{L_1 L_2 C_1 C_2 R_D C_D s^5 + L_1 L_2 C_2 (C_1 + C_D) s^4 + L_1 R_D C_D (C_1 + C_2) s^3 + \cdots}$$

$$\cdots \frac{\cdots}{\cdots + L_1(C_1 + C_2 + C_D)s^2 + L_2 C_2 R_D C_D s^3 + L_2 C_2 s^2 + R_D C_D s + 1} \, ;$$

$$G_1(s) = \frac{(L_2 C_2 s^2 + 1) \cdot (R_D C_D s + 1)}{L_1 L_2 C_1 C_2 R_D C_D s^5 + L_1 L_2 C_2 (C_1 + C_D) s^4 + R_D C_D [L_1 (C_1 + C_2) + L_2 C_2] s^3 + \cdots}$$

$$\cdots \frac{\cdots}{\cdots + [L_1(C_1 + C_2 + C_D) + L_2 C_2] s^2 + R_D C_D s + 1} \, ;$$

$$G_2(s) = \frac{\frac{1}{C_2 s}}{L_2 s + \frac{1}{C_2 s}} = \frac{1}{L_2 C_2 s^2 + 1} \, ;$$

$$G(s) = G_1(s) \cdot G_2(s);$$

$$G(s) = \frac{R_D C_D s + 1}{L_1 L_2 C_1 C_2 R_D C_D s^5 + L_1 L_2 C_2 (C_1 + C_D) s^4 + R_D C_D [L_1 (C_1 + C_2) + L_2 C_2] s^3 + \cdots} \quad (13)$$

$$\cdots \frac{\cdots}{\cdots + [L_1(C_1 + C_2 + C_D) + L_2 C_2] s^2 + R_D C_D s + 1}.$$

Note that the *s* terms in the numerator and the denominator are equal. Therefore we can write Eq. (13) as follows:

$$G(s) = \frac{k_1 s + 1}{k_5 s^5 + k_4 s^4 + k_3 s^3 + k_2 s^2 + k_1 s + 1} \quad (14a)$$

$$\text{with} \quad k_1 = R_D C_D, \quad (14b)$$

$$k_2 = L_1(C_1 + C_2 + C_D) + L_2 C_2, \quad (14c)$$

$$k_3 = R_D C_D (L_1 C_1 + L_2 C_2 + L_1 C_2), \quad (14d)$$

$$k_4 = L_1 L_2 C_2 (C_1 + C_D), \quad (14e)$$

$$k_5 = L_1 L_2 C_1 C_2 C_D R_D. \quad (14f)$$

*G*(*s*) can be expressed as the combination of a fifth-order PT and a first-order PD. Let us have a closer look at the fifth-order PT, which is the denominator part of *G*(*s*). This can be split in two second-order PTs and one first-order PT, all connected in series. It can be expressed in its normalized form as

$$G_{PT}(s) = \frac{1}{(1 + a_1 \frac{s}{\omega_0}) \cdot (1 + a_2 \frac{s}{\omega_0} + b_2 \frac{s^2}{\omega_0^2}) \cdot (1 + a_3 \frac{s}{\omega_0} + b_3 \frac{s^2}{\omega_0^2})}. \quad (15)$$

The transfer function of such a combination of first- and second-order PTs can be optimized according to different methods [1], which results in particular values for $a_i$ and $b_i$. The commonly used optimization methods with their corresponding coefficients are given in Table 5.

**Table 5:** Coefficients for a fifth-order PT for different optimization methods

| Method | $a_1$ | $a_2$ | $b_2$ | $a_3$ | $b_3$ |
|---|---|---|---|---|---|
| Butterworth | 1.0000 | 1.6180 | 1.0000 | 0.6180 | 1.0000 |
| Bessel | 0.6656 | 1.1402 | 0.4128 | 0.6216 | 0.3245 |
| Critical damping | 0.3856 | 0.7712 | 0.1487 | 0.7712 | 0.1487 |

By expanding Eq. (15) we get

$$G_{PT}(s) = \frac{1}{(1 + \frac{a_2}{\omega_0}s + \frac{b_2}{\omega_0^2}s^2 + \frac{a_1}{\omega_0}s + \frac{a_1 a_2}{\omega_0^2}s^2 + \frac{a_1 b_2}{\omega_0^3}s^3) \cdot (1 + \frac{a_3}{\omega_0}s + \frac{b_3}{\omega_0^2}s^2)}$$

$$= \frac{1}{1 + \frac{a_3}{\omega_0}s + \frac{b_3}{\omega_0^2}s^2 + \frac{a_2}{\omega_0}s + \frac{a_2 a_3}{\omega_0^2}s^2 + \frac{a_2 b_3}{\omega_0^3}s^3 + \frac{b_2}{\omega_0^2}s^2 + \frac{a_3 b_2}{\omega_0^3}s^3 + \cdots}$$

$$\cdots \overline{\cdots + \frac{b_2 b_3}{\omega_0^4}s^4 + \frac{a_1}{\omega_0}s + \frac{a_1 a_3}{\omega_0^2}s^2 + \frac{a_1 b_3}{\omega_0^3}s^3 + \frac{a_1 a_2}{\omega_0^2}s^2 + \frac{a_1 a_2 a_3}{\omega_0^3}s^3 + \cdots} \cdots$$

$$\cdots \overline{\cdots + \frac{a_1 a_2 b_3}{\omega_0^4}s^4 + \frac{a_1 b_2}{\omega_0^3}s^3 + \frac{a_1 a_3 b_2}{\omega_0^4}s^4 + \frac{a_1 b_2 b_3}{\omega_0^5}s^5};$$

$$G_{PT}(s) = \frac{1}{\frac{a_1 b_2 b_3}{\omega_0^5}s^5 + \frac{(b_2 b_3 + a_1 a_2 b_3 + a_1 a_3 b_2)}{\omega_0^4}s^4 + \cdots} \quad (16)$$

$$\cdots \overline{\cdots + \frac{(a_2 b_3 + a_3 b_2 + a_1 b_3 + a_1 a_2 a_3 + a_1 b_2)}{\omega_0^3}s^3 + \cdots} \cdots$$

$$\cdots \overline{\cdots + \frac{(b_3 + a_2 a_3 + b_2 + a_1 a_3 + a_1 a_2)}{\omega_0^2}s^2 + \frac{(a_1 + a_2 + a_3)}{\omega_0}s + 1}.$$

By comparing the coefficients with Eq. (14a) we get

$$k_1 = R_D C_D = \frac{a_1 + a_2 + a_3}{\omega_0}; \tag{17a}$$

$$k_2 = L_1(C_1 + C_2 + C_D) + L_2 C_2 = \frac{b_3 + a_2 a_3 + b_2 + a_1 a_3 + a_1 a_2}{\omega_0^2}; \tag{17b}$$

$$k_3 = R_D C_D (L_1 C_1 + L_2 C_2 + L_1 C_2) = \frac{a_2 b_3 + a_3 b_2 + a_1 b_3 + a_1 a_2 a_3 + a_1 b_2}{\omega_0^3}; \tag{17c}$$

$$k_4 = L_1 L_2 C_2 (C_1 + C_D) = \frac{b_2 b_3 + a_1 a_2 b_3 + a_1 a_3 b_2}{\omega_0^4}; \tag{17d}$$

$$k_5 = L_1 L_2 C_1 C_2 C_D R_D = \frac{a_1 b_2 b_3}{\omega_0^5}. \tag{17e}$$

For a given optimization method, the five independent Eqs. (17a)–(17e) contain seven unknowns ($L_1$, $L_2$, $C_1$, $C_2$, $R_D$, $C_D$, and $\omega_0$). Therefore, we have the choice to select two of them and the remaining five depend on that selection; here we preselect $\omega_0$ and $L_1$. Refer to Section 2 for the evaluation of $L_1$.

Selection of the cut-off angular frequency $\omega_0$: For a given angular frequency $\omega_B$ well in the blocking area of the filter ($\omega_B \gg \omega_0$) we can define the desired attenuation $G_B$. In the blocking area the highest-order terms of both the numerator and the denominator in Eq. (13) dominate, therefore it can be simplified to

$$G_B = \frac{R_D C_D s}{L_1 L_2 C_1 C_2 R_D C_D s^5} = \frac{\frac{a_1 + a_2 + a_3}{\omega_0} s}{\frac{a_1 b_2 b_3}{\omega_0^5} s^5} = \frac{a_1 + a_2 + a_3}{a_1 b_2 b_3} \cdot \frac{\omega_0^4}{s^4} = \frac{a_1 + a_2 + a_3}{a_1 b_2 b_3} \cdot \frac{\omega_0^4}{\omega_B^4},$$

$$\omega_0 = \omega_B \cdot \sqrt[4]{\frac{G_B \cdot a_1 b_2 b_3}{a_1 + a_2 + a_3}}. \tag{18}$$

The cut-off angular frequency $\omega_0$ of the filter depends only on the required attenuation and on the selected optimization method. The equation system (17a)–(17e) has to be solved for $L_2$, $C_1$, $C_2$, $R_D$, and $C_D$.

By solving Eq. (17a) for $C_D$ and substituting $C_D$ into Eqs. (17b)–(17e) we reduce the system to four equations:

$$k_2 = L_1 \left( C_1 + C_2 + \frac{k_1}{R_D} \right) + L_2 C_2; \tag{19a}$$

$$k_3 = k_1 (L_1 C_1 + L_2 C_2 + L_1 C_2); \tag{19b}$$

$$k_4 = L_1 L_2 C_2 \left( C_1 + \frac{k_1}{R_D} \right); \tag{19c}$$

$$k_5 = k_1 L_1 L_2 C_1 C_2. \tag{19d}$$

By dividing Eq. (19c) by Eq. (19d) we get

$$\frac{k_4}{k_5} = \frac{L_1 L_2 C_2 \left( C_1 + \frac{k_1}{R_D} \right)}{k_1 L_1 L_2 C_1 C_2} = \frac{\left( C_1 + \frac{k_1}{R_D} \right)}{k_1 C_1} = \frac{1}{k_1} + \frac{1}{C_1 R_D},$$

$$k_1 k_4 R_D C_1 - k_5 R_D C_1 = k_1 k_5,$$

$$R_D = \frac{k_1 k_5}{C_1(k_1 k_4 - k_5)}. \tag{20}$$

By substituting Eq. (20) in Eq. (19a) we reduce the system further to three equations:

$$k_2 = L_1\left(C_1 + C_2 + \frac{k_1 C_1(k_1 k_4 - k_5)}{k_1 k_5}\right) + L_2 C_2,$$

$$= L_1\left(C_2 + C_1(1 + \frac{k_1 k_4}{k_5} - 1)\right) + L_2 C_2 = L_1 C_2 + L_1 C_1 \frac{k_1 k_4}{k_5} + L_2 C_2,$$

$$= (L_1 + L_2)C_2 + L_1 C_1 \frac{k_1 k_4}{k_5}; \tag{21a}$$

$$k_3 = k_1(L_1 C_1 + L_2 C_2 + L_1 C_2); \tag{21b}$$

$$k_5 = k_1 L_1 L_2 C_1 C_2. \tag{21c}$$

By solving Eq. (21c) for $C_1$ and substituting $C_1$ in Eqs. (21a) and (21b), we reduce the system further to two equations:

$$k_2 = (L_1 + L_2)C_2 + L_1 \cdot \frac{k_5}{k_1 L_1 L_2 C_2} \cdot \frac{k_1 k_4}{k_5} = C_2(L_1 + L_2) + \frac{k_4}{L_2 C_2}, \tag{22a}$$

$$k_3 = k_1\left(L_1 \cdot \frac{k_5}{k_1 L_1 L_2 C_2} + L_2 C_2 + L_1 C_2\right) = k_1 C_2(L_1 + L_2) + \frac{k_5}{L_2 C_2}. \tag{22b}$$

By subtracting Eq. (22b) from Eq. (22a) we get

$$k_2 - \frac{k_3}{k_1} = \frac{k_1 k_2 - k_3}{k_1} = C_2(L_1 + L_2) + \frac{k_4}{L_2 C_2} - C_2(L_1 + L_2) - \frac{k_5}{k_1 L_2 C_2} = \frac{k_1 k_4 - k_5}{k_1 L_2 C_2};$$

$$C_2 = \frac{k_1 k_4 - k_5}{L_2(k_1 k_2 - k_3)}. \tag{23}$$

By substituting Eq. (23) in Eq. (22a) we get

$$k_2 = \frac{(k_1 k_4 - k_5) \cdot (L_1 + L_2)}{L_2(k_1 k_2 - k_3)} + \frac{k_4 L_2(k_1 k_2 - k_3)}{L_2(k_1 k_4 - k_5)};$$

$$\frac{(k_1 k_4 - k_5) \cdot (L_1 + L_2)}{L_2(k_1 k_2 - k_3)} = k_2 - \frac{k_4(k_1 k_2 - k_3)}{k_1 k_4 - k_5} = \frac{k_1 k_2 k_4 - k_2 k_5 - k_1 k_2 k_4 + k_3 k_4}{k_1 k_4 - k_5}$$

$$= \frac{k_3 k_4 - k_2 k_5}{k_1 k_4 - k_5};$$

$$\frac{L_1 + L_2}{L_2} = \frac{L_1}{L_2} + 1 = \frac{(k_3 k_4 - k_2 k_5) \cdot (k_1 k_2 - k_3)}{(k_1 k_4 - k_5)^2};$$

$$L_1 = L_2\left[\frac{(k_3 k_4 - k_2 k_5) \cdot (k_1 k_2 - k_3)}{(k_1 k_4 - k_5)^2} - 1\right];$$

$$L_2 = \frac{L_1}{\frac{(k_3 k_4 - k_2 k_5) \cdot (k_1 k_2 - k_3)}{(k_1 k_4 - k_5)^2} - 1}. \tag{24}$$

Example 3: Design a fourth-order filter for a buck converter with a d.c.-link voltage of 120 V, a switching frequency of 20 kHz, and a maximum output current of 500 A; refer to Fig. 10. The ripple current in $L_1$ should not exceed 50 A peak to peak. The filter should have an attenuation of 0.004 at the switching frequency. Design the filter for all three optimization methods and compare the results. In order to enable a comparison between different filter structures (refer to Section 4), the same parameters are given as in Example 2.

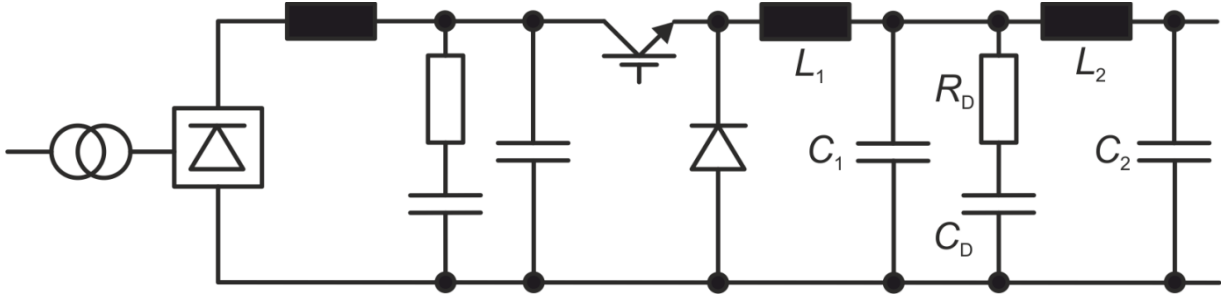

**Fig. 10:** Output filter for a buck converter

Select $L_1$ in order to meet the ripple requirement by using Eq. (8a):

$$L_1 = \frac{V_{\text{dc}} \cdot 0.25}{f_s \cdot \Delta I_{L1}} = \frac{120 \text{ V} \cdot 0.25}{20 \text{ kHz} \cdot 50 \text{ A}} = 30 \ \mu\text{H}.$$

Select $\omega_0$ in order to meet the attenuation requirement by using Eq. (18):

$$\omega_0 = \omega_B \cdot \sqrt[4]{\frac{G_B \cdot a_1 b_2 b_3}{a_1 + a_2 + a_3}} = 2 \cdot \pi \cdot 20 \text{ kHz} \cdot \sqrt[4]{\frac{0.004 \cdot a_1 b_2 b_3}{a_1 + a_2 + a_3}}.$$

Calculate the remaining filter components by using Eqs. (24), (23), (21c), (20), and (17a) in that order. The results are listed in Table 6. Depending on the selected optimization method, $L_2$ is approximately either double, equal to, or half the size of $L_1$. This leads to a simplified design as $L_1$ and $L_2$ can be realized with either two or three identical chokes.

**Table 6:** Results for Example 3

| Parameter | Butterworth | Bessel | Critical damping |
|---|---|---|---|
| $\omega_0$ | 23,600 s$^{-1}$ | 13,800 s$^{-1}$ | 8200 s$^{-1}$ |
| $f_0$ | 3.75 kHz | 2.20 kHz | 1.30 kHz |
| $L_1$ | 30 $\mu$H | 30 $\mu$H | 30 $\mu$H |
| $L_2$ | 57 $\mu$H | 31 $\mu$H | 17 $\mu$H |
| $C_1$ | 23 $\mu$F | 24 $\mu$F | 25 $\mu$F |
| $C_2$ | 26 $\mu$F | 44 $\mu$F | 80 $\mu$F |
| $C_D$ | 217 $\mu$F | 342 $\mu$F | 597 $\mu$F |
| $R_D$ | 0.63 $\Omega$ | 0.51 $\Omega$ | 0.40 $\Omega$ |

Figure 11 shows the Bode plots of the three filter designs according to Example 3. The attenuation at 20 kHz is −48 dB, which corresponds to a factor of 0.004, as required. Compared to the second-order filter (see Fig. 8) the resonance amplitudes are slightly higher and the same trade-off between small capacitor values and low resonance amplitudes applies.

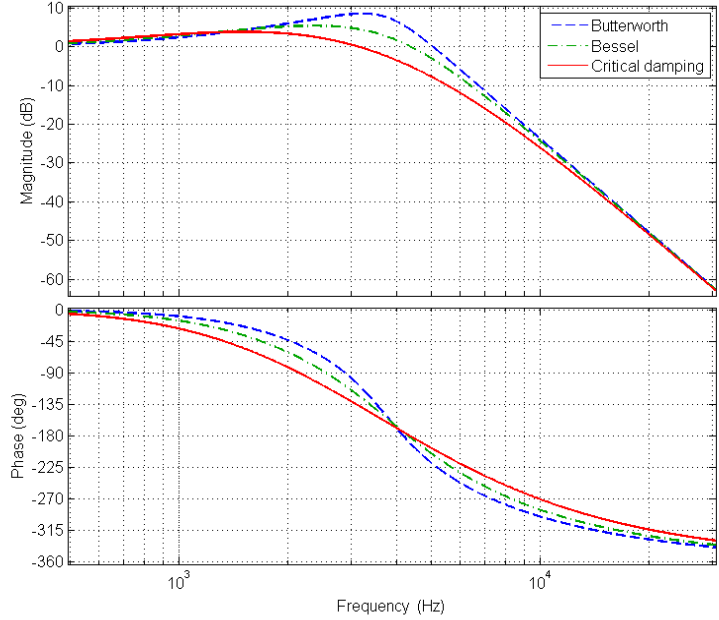

**Fig. 11:** Bode plots for the filter designs according to Example 3

### 3.2 Fourth-order low-pass filter with damping circuit in the second *LC* stage

In this section a low-pass filter according to Fig. 12 is outlined. The derivation is similar to the one in Section 3.1. For the sake of completeness it is repeated in detail.

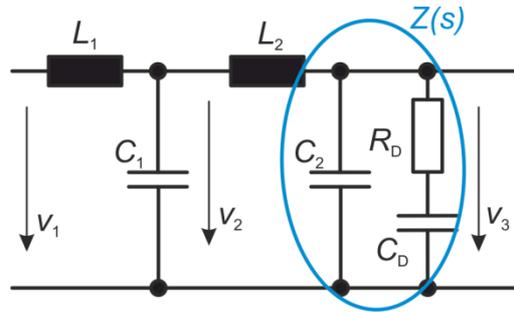

**Fig. 12:** Fourth-order low-pass filter with *RC* damping in the second *LC* stage

The complex impedance *Z*(*s*) is given by

$$Z(s) = \frac{1}{C_2 s + \dfrac{1}{R_\mathrm{D} + \dfrac{1}{C_\mathrm{D} s}}} = \frac{1}{C_2 s + \dfrac{C_\mathrm{D} s}{R_\mathrm{D} C_\mathrm{D} s + 1}} = \frac{R_\mathrm{D} C_\mathrm{D} s + 1}{C_2 R_\mathrm{D} C_\mathrm{D} s^2 + (C_2 + C_\mathrm{D}) s}. \qquad (25)$$

The transfer function can be split into two partial transfer functions $G_1(s)$ and $G_2(s)$:

$$G(s) = G_1(s) \cdot G_2(s) = \frac{v_2(s)}{v_1(s)} \cdot \frac{v_3(s)}{v_2(s)} = \frac{v_3(s)}{v_1(s)};$$

$$G_1(s) = \frac{C_1 s + \frac{1}{L_2 s + Z(s)}}{L_1 s + \frac{1}{C_1 s + \frac{1}{L_2 s + Z(s)}}} = \frac{\frac{L_2 s + Z(s)}{C_1 L_2 s^2 + C_1 Z(s) s + 1}}{L_1 s + \frac{L_2 s + Z(s)}{C_1 L_2 s^2 + C_1 Z(s) s + 1}}$$

$$= \frac{L_2 s + Z(s)}{L_1 L_2 C_1 s^3 + L_1 C_1 Z(s) s^2 + (L_1 + L_2) s + Z(s)};$$

$$G_2(s) = \frac{Z(s)}{L_2 s + Z(s)};$$

$$G(s) = G_1(s) \cdot G_2(s) = \frac{Z(s)}{L_1 L_2 C_1 s^3 + L_1 C_1 Z(s) s^2 + (L_1 + L_2) s + Z(s)}$$

$$= \frac{\frac{R_D C_D s + 1}{C_2 R_D C_D s^2 + (C_2 + C_D) s}}{L_1 L_2 C_1 s^3 + \frac{L_1 C_1 R_D C_D s + L_1 C_1}{C_2 R_D C_D s^2 + (C_2 + C_D) s} s^2 + (L_1 + L_2) s + \frac{R_D C_D s + 1}{C_2 R_D C_D s^2 + (C_2 + C_D) s}}$$

$$= \frac{R_D C_D s + 1}{L_1 L_2 C_1 C_2 R_D C_D s^5 + (C_2 + C_D) L_1 L_2 C_1 s^4 + L_1 C_1 R_D C_D s^3 + L_1 C_1 s^2 + \cdots} \quad (26)$$

$$\overline{\cdots + (L_1 + L_2) C_2 R_D C_D s^3 + (L_1 + L_2)(C_2 + C_D) s^2 + R_D C_D s + 1};$$

$$G(s) = \frac{R_D C_D s + 1}{L_1 L_2 C_1 C_2 R_D C_D s^5 + (C_2 + C_D) L_1 L_2 C_1 s^4 + \cdots}$$

$$\overline{\cdots + [L_1 C_1 R_D C_D + (L_1 + L_2) C_2 R_D C_D] s^3 + [L_1 C_1 + (L_1 + L_2)(C_2 + C_D)] s^2 + R_D C_D s + 1}.$$

Note that the $s$ terms in the numerator and the denominator are equal. Therefore we can write Eq. (26) as

$$G(s) = \frac{k_1 s + 1}{k_5 s^5 + k_4 s^4 + k_3 s^3 + k_2 s^2 + k_1 s + 1}, \tag{27a}$$

$$\text{with} \quad k_1 = R_D C_D; \tag{27b}$$

$$k_2 = L_1(C_1 + C_2 + C_D) + L_2(C_2 + C_D); \tag{27c}$$

$$k_3 = R_D C_D (L_1 C_1 + L_2 C_2 + L_1 C_2); \tag{27d}$$

$$k_4 = L_1 L_2 C_1 (C_2 + C_D); \tag{27e}$$

$$k_5 = L_1 L_2 C_1 C_2 C_D R_D. \tag{27f}$$

$G(s)$ can be expressed as the combination of a fifth-order PT and a first-order PD. Let us take a closer look at the fifth-order PT, which is the denominator part of $G(s)$. This can be split in two second-order PTs and one first-order PT, all connected in series. It can be expressed in its normalized form as

$$G_{\text{PT}}(s) = \frac{1}{(1 + a_1 \frac{s}{\omega_0}) \cdot (1 + a_2 \frac{s}{\omega_0} + b_2 \frac{s^2}{\omega_0^2}) \cdot (1 + a_3 \frac{s}{\omega_0} + b_3 \frac{s^2}{\omega_0^2})}. \tag{28}$$

The transfer function of such a combination of first- and second-order PTs can be optimized according to different methods [1], which results in particular values for $a_i$ and $b_i$. The commonly used optimization methods with their corresponding coefficients are given in Table 7.

**Table 7:** Coefficients for a fifth-order PT for different optimization methods

| Method | $a_1$ | $a_2$ | $b_2$ | $a_3$ | $b_3$ |
|---|---|---|---|---|---|
| Butterworth | 1.0000 | 1.6180 | 1.0000 | 0.6180 | 1.0000 |
| Bessel | 0.6656 | 1.1402 | 0.4128 | 0.6216 | 0.3245 |
| Critical damping | 0.3856 | 0.7712 | 0.1487 | 0.7712 | 0.1487 |

By expanding Eq. (28) we get

$$G_{\text{PT}}(s) = \frac{1}{(1 + \frac{a_2}{\omega_0}s + \frac{b_2}{\omega_0^2}s^2 + \frac{a_1}{\omega_0}s + \frac{a_1 a_2}{\omega_0^2}s^2 + \frac{a_1 b_2}{\omega_0^3}s^3) \cdot (1 + \frac{a_3}{\omega_0}s + \frac{b_3}{\omega_0^2}s^2)}$$

$$= \frac{1}{1 + \frac{a_3}{\omega_0}s + \frac{b_3}{\omega_0^2}s^2 + \frac{a_2}{\omega_0}s + \frac{a_2 a_3}{\omega_0^2}s^2 + \frac{a_2 b_3}{\omega_0^3}s^3 + \frac{b_2}{\omega_0^2}s^2 + \frac{a_3 b_2}{\omega_0^3}s^3 + \cdots}$$

$$\cdots + \frac{b_2 b_3}{\omega_0^4}s^4 + \frac{a_1}{\omega_0}s + \frac{a_1 a_3}{\omega_0^2}s^2 + \frac{a_1 b_3}{\omega_0^3}s^3 + \frac{a_1 a_2}{\omega_0^2}s^2 + \frac{a_1 a_2 a_3}{\omega_0^3}s^3 + \cdots$$

$$\cdots + \frac{a_1 a_2 b_3}{\omega_0^4}s^4 + \frac{a_1 b_2}{\omega_0^3}s^3 + \frac{a_1 a_3 b_2}{\omega_0^4}s^4 + \frac{a_1 b_2 b_3}{\omega_0^5}s^5;$$

$$G_{\text{PT}}(s) = \frac{1}{\frac{a_1 b_2 b_3}{\omega_0^5}s^5 + \frac{(b_2 b_3 + a_1 a_2 b_3 + a_1 a_3 b_2)}{\omega_0^4}s^4 + \cdots} \tag{29}$$

$$\cdots + \frac{(a_2 b_3 + a_3 b_2 + a_1 b_3 + a_1 a_2 a_3 + a_1 b_2)}{\omega_0^3}s^3 + \cdots$$

$$\cdots + \frac{(b_3 + a_2 a_3 + b_2 + a_1 a_3 + a_1 a_2)}{\omega_0^2}s^2 + \frac{(a_1 + a_2 + a_3)}{\omega_0}s + 1.$$

By comparing the coefficients with Eq. (27a) we get

$$k_1 = R_D C_D = \frac{a_1 + a_2 + a_3}{\omega_0}; \tag{30a}$$

$$k_2 = L_1(C_1 + C_2 + C_D) + L_2(C_2 + C_D) = \frac{b_3 + a_2 a_3 + b_2 + a_1 a_3 + a_1 a_2}{\omega_0^2}; \tag{30b}$$

$$k_3 = R_D C_D (L_1 C_1 + L_2 C_2 + L_1 C_2) = \frac{a_2 b_3 + a_3 b_2 + a_1 b_3 + a_1 a_2 a_3 + a_1 b_2}{\omega_0^3}; \tag{30c}$$

$$k_4 = L_1 L_2 C_1 (C_2 + C_D) = \frac{b_2 b_3 + a_1 a_2 b_3 + a_1 a_3 b_2}{\omega_0^4}; \tag{30d}$$

$$k_5 = L_1 L_2 C_1 C_2 C_D R_D = \frac{a_1 b_2 b_3}{\omega_0^5}. \tag{30e}$$

For a given optimization method, the five independent Eqs. (30a)–(30e) contain seven unknowns ($L_1$, $L_2$, $C_1$, $C_2$, $R_D$, $C_D$, and $\omega_0$). Therefore we have the choice to select two of them and the remaining five depend on that selection; here we preselect $\omega_0$ and $L_1$. Refer to Section 2 for the evaluation of $L_1$.

Selection of the cut-off angular frequency $\omega_0$: For a given angular frequency $\omega_B$ well in the blocking area of the filter ($\omega_B \gg \omega_0$) we can define the desired attenuation $G_B$. In the blocking area the highest-order terms of both the numerator and the denominator in Eq. (26) dominate, therefore it can be simplified to

$$G_B = \frac{R_D C_D s}{L_1 L_2 C_1 C_2 R_D C_D s^5} = \frac{\frac{a_1 + a_2 + a_3}{\omega_0} s}{\frac{a_1 b_2 b_3}{\omega_0^5} s^5} = \frac{a_1 + a_2 + a_3}{a_1 b_2 b_3} \cdot \frac{\omega_0^4}{s^4} = \frac{a_1 + a_2 + a_3}{a_1 b_2 b_3} \cdot \frac{\omega_0^4}{\omega_B^4};$$

$$\omega_0 = \omega_B \cdot \sqrt[4]{\frac{G_B \cdot a_1 b_2 b_3}{a_1 + a_2 + a_3}}. \tag{31}$$

The cut-off angular frequency $\omega_0$ of the filter depends only on the required attenuation and on the selected optimization method. The equation system (30a)–(30e) has to be solved for $L_2$, $C_1$, $C_2$, $R_D$, and $C_D$.

By solving Eq. (30a) for $C_D$ and substituting $C_D$ in Eqs. (30b)–(30e) we reduce the system to four equations:

$$k_2 = L_1 \left( C_1 + C_2 + \frac{k_1}{R_D} \right) + L_2 \left( C_2 + \frac{k_1}{R_D} \right); \tag{32a}$$

$$k_3 = k_1 (L_1 C_1 + L_2 C_2 + L_1 C_2); \tag{32b}$$

$$k_4 = L_1 L_2 C_1 \left( C_2 + \frac{k_1}{R_D} \right); \tag{32c}$$

$$k_5 = k_1 L_1 L_2 C_1 C_2. \tag{32d}$$

By dividing Eq. (32c) by Eq. (32d) we get:

$$\frac{k_4}{k_5} = \frac{L_1 L_2 C_1 \left(C_2 + \frac{k_1}{R_D}\right)}{k_1 L_1 L_2 C_1 C_2} = \frac{\left(C_2 + \frac{k_1}{R_D}\right)}{k_1 C_2};$$

$$k_1 k_4 C_2 - k_5 C_2 = \frac{k_1 k_5}{R_D};$$

$$R_D = \frac{k_1 k_5}{C_2 (k_1 k_4 - k_5)}. \tag{33}$$

By substituting Eq. (33) in Eq. (32a) we reduce the system further to three equations:

$$k_2 = L_1 \left(C_1 + C_2 + \frac{k_1 C_2 (k_1 k_4 - k_5)}{k_1 k_5}\right) + L_2 \left(C_2 + \frac{k_1 C_2 (k_1 k_4 - k_5)}{k_1 k_5}\right)$$

$$= L_1 \left(C_1 + C_2 (1 + \frac{k_1 k_4}{k_5} - 1)\right) + L_2 \left(C_2 (1 + \frac{k_1 k_4}{k_5} - 1)\right) = L_1 C_1 + \frac{k_1 k_4}{k_5} L_1 C_2 + \frac{k_1 k_4}{k_5} L_2 C_2$$

$$= L_1 C_1 + C_2 (L_1 + L_2) \frac{k_1 k_4}{k_5}; \tag{34a}$$

$$k_3 = k_1 (L_1 C_1 + L_2 C_2 + L_1 C_2); \tag{34b}$$

$$k_5 = k_1 L_1 L_2 C_1 C_2. \tag{34c}$$

By solving Eq. (34c) for $C_1$ and substituting $C_1$ in Eqs. (34a) and (34b) we reduce the system further to two equations:

$$k_2 = \frac{k_5}{k_1 L_2 C_2} + C_2 (L_1 + L_2) \frac{k_1 k_4}{k_5}; \tag{35a}$$

$$\frac{k_3}{k_1} = \frac{k_5}{k_1 L_2 C_2} + C_2 (L_1 + L_2). \tag{35b}$$

By subtracting Eq. (35b) from Eq. (35a) we get

$$k_2 - \frac{k_3}{k_1} = C_2 (L_1 + L_2) \left(\frac{k_1 k_4}{k_5} - 1\right) = \frac{k_1 k_2 - k_3}{k_1} = \frac{k_1 k_4 - k_5}{k_5} C_2 (L_1 + L_2);$$

$$C_2 = \frac{k_5 (k_1 k_2 - k_3)}{k_1 (k_1 k_4 - k_5)(L_1 + L_2)}. \tag{36}$$

By substituting Eq. (36) in Eq. (35a) we get

$$k_2 = \frac{k_5 k_1 (k_1 k_4 - k_5)(L_1 + L_2)}{k_1 k_5 (k_1 k_2 - k_3) L_2} + \frac{k_1 k_4 k_5 (k_1 k_2 - k_3)(L_1 + L_2)}{k_1 k_5 (k_1 k_4 - k_5)(L_1 + L_2)}$$

$$= \frac{(k_1 k_4 - k_5)(L_1 + L_2)}{(k_1 k_2 - k_3) L_2} + \frac{k_4 (k_1 k_2 - k_3)}{(k_1 k_4 - k_5)};$$

$$\frac{(k_1k_4 - k_5)(L_1 + L_2)}{(k_1k_2 - k_3)L_2} = k_2 - \frac{k_4(k_1k_2 - k_3)}{k_1k_4 - k_5} = \frac{k_1k_2k_4 - k_2k_5 - k_1k_2k_4 + k_3k_4}{k_1k_4 - k_5}$$
$$= \frac{k_3k_4 - k_2k_5}{k_1k_4 - k_5};$$

$$\frac{L_1 + L_2}{L_2} = \frac{L_1}{L_2} + 1 = \frac{(k_3k_4 - k_2k_5)(k_1k_2 - k_3)}{(k_1k_4 - k_5)^2};$$

$$L_1 = L_2 \left[ \frac{(k_3k_4 - k_2k_5)(k_1k_2 - k_3)}{(k_1k_4 - k_5)^2} - 1 \right];$$

$$L_2 = \frac{L_1}{\frac{(k_3k_4 - k_2k_5)(k_1k_2 - k_3)}{(k_1k_4 - k_5)^2} - 1}. \tag{37}$$

Example 4: Design a fourth-order filter for a buck converter with a d.c.-link voltage of 120 V, a switching frequency of 20 kHz, and a maximum output current of 500 A; refer to Fig. 13. The ripple current in $L_1$ should not exceed 50 A peak to peak. The filter should have an attenuation of 0.004 at the switching frequency. Design the filter for all three optimization methods and compare the results. In order to enable a comparison between different filter structures (refer to Section 4), the same parameters are given as in Examples 2 and 3.

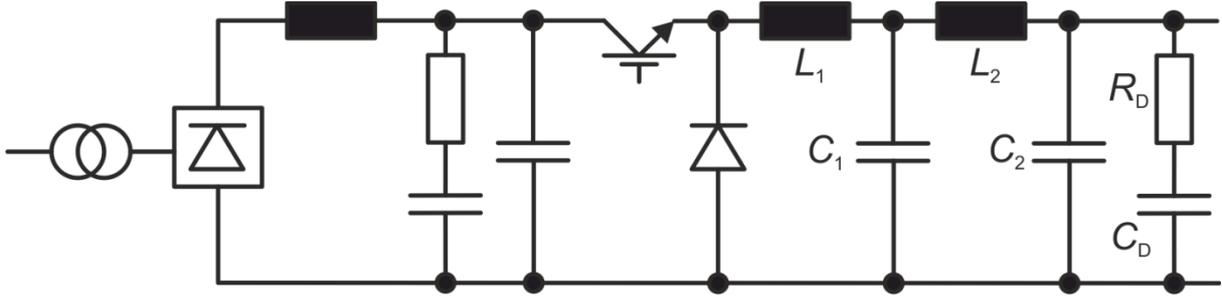

**Fig. 13:** Output filter for a buck converter

Select $L_1$ in order to meet the ripple requirement by using Eq. (8a):

$$L_1 = \frac{V_{dc} \cdot 0.25}{f_s \cdot \Delta I_{L1}} = \frac{120 \text{ V} \cdot 0.25}{20 \text{ kHz} \cdot 50 \text{ A}} = 30 \text{ μH}.$$

Select $\omega_0$ in order to meet the attenuation requirement by using Eq. (31)

$$\omega_0 = \omega_B \cdot \sqrt[4]{\frac{G_B \cdot a_1 b_2 b_3}{a_1 + a_2 + a_3}} = 2 \cdot \pi \cdot 20 \text{ kHz} \cdot \sqrt[4]{\frac{0.004 \cdot a_1 b_2 b_3}{a_1 + a_2 + a_3}}.$$

Calculate the remaining filter components by using Eqs. (37), (36), (34c), (33) and (30a) in that order. The results are listed in Table 8. Depending on the selected optimization method, $L_2$ is approximately either double, equal to, or half the size of $L_1$. This leads to a simplified design as $L_1$ and $L_2$ can be realized with either two or three identical chokes.

**Table 8:** Results for Example 4

| Parameter | Butterworth | Bessel | Critical damping |
|---|---|---|---|
| $\omega_0$ | 23,600 s$^{-1}$ | 13,800 s$^{-1}$ | 8200 s$^{-1}$ |
| $f_0$ | 3.75 kHz | 2.20 kHz | 1.30 kHz |
| $L_1$ | 30 $\mu$H | 30 $\mu$H | 30 $\mu$H |
| $L_2$ | 57 $\mu$H | 31 $\mu$H | 17 $\mu$H |
| $C_1$ | 74 $\mu$F | 90 $\mu$F | 124 $\mu$F |
| $C_2$ | 7.9 $\mu$F | 12 $\mu$F | 16 $\mu$F |
| $C_D$ | 75 $\mu$F | 168 $\mu$F | 382 $\mu$F |
| $R_D$ | 1.83 $\Omega$ | 1.05 $\Omega$ | 0.62 $\Omega$ |

Figure 14 shows the Bode plots of the three filter designs according to Example 4. They are exactly the same as for Example 3. The attenuation at 20 kHz is −48 dB, which corresponds to a factor of 0.004 as required. Compared to the second-order filter (see Fig. 8) the resonance amplitudes are slightly higher and the same trade-off between small capacitor values and low resonance amplitudes applies.

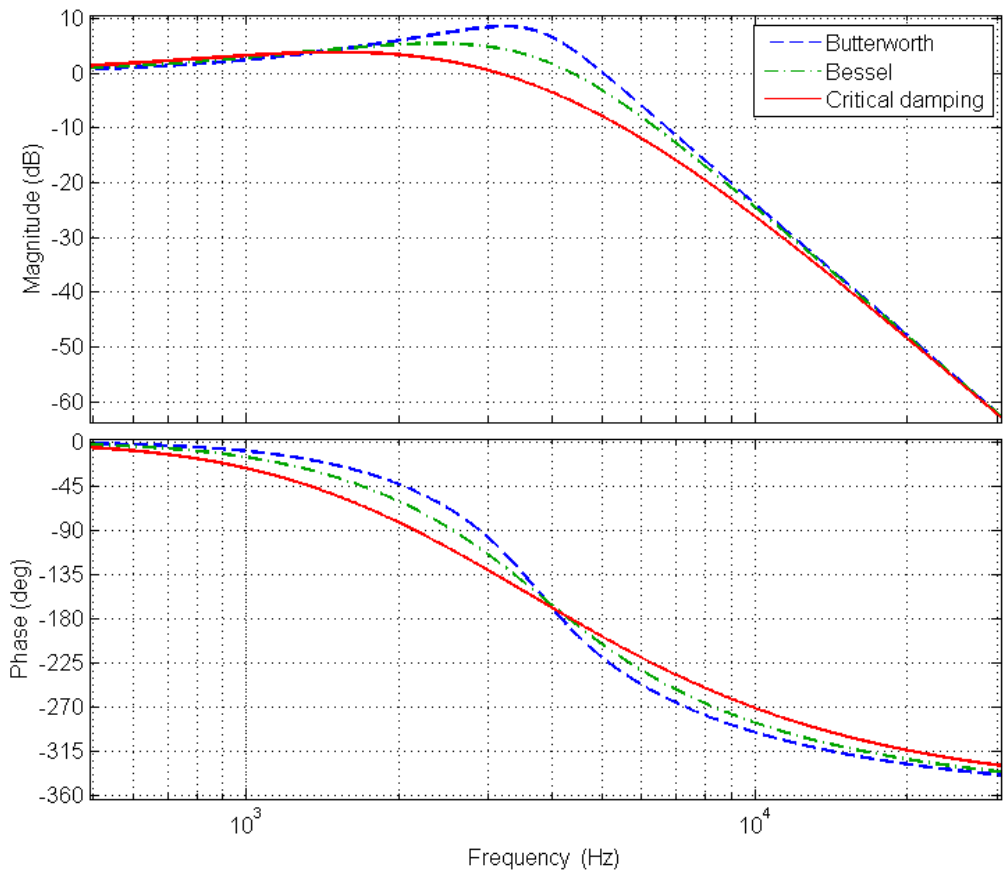

**Fig. 14:** Bode plots for the filter designs according to Example 4

## 4 Comparison of different filter designs

In Examples 2–4 we have designed three filters with different structures but with the same performance (attenuation factor of 0.004 at 20 kHz). This allows a direct comparison of the three filter structures. Table 9 summarizes the results for the Bessel optimization.

**Table 9:** Results for Bessel optimization Examples 2–4

| Parameter | Example 2 | Example 2 $L_1 = 100\ \mu H$ | Example 2 $L_1 = 100\ \mu H$ $G_B = 0.01$ | Example 3 | Example 4 |
|---|---|---|---|---|---|
| $L_1$ | 30 $\mu$H | 100 $\mu$H | 100 $\mu$H | 30 $\mu$H | 30 $\mu$H |
| $L_2$ | | | | 31 $\mu$H | 31 $\mu$H |
| $C_1$ | 528 $\mu$F | 158 $\mu$F | 63 $\mu$F | 24 $\mu$F | 90 $\mu$F |
| $C_2$ | | | | 44 $\mu$F | 12 $\mu$F |
| $C_D$ | 2640 $\mu$F | 790 $\mu$F | 320 $\mu$F | 342 $\mu$F | 168 $\mu$F |
| $C_1 + C_2 + C_D$ | 3168 $\mu$F | 948 $\mu$F | 383 $\mu$F | 410 $\mu$F | 270 $\mu$F |
| $R_D$ | 0.18 $\Omega$ | 0.62 $\Omega$ | 0.98 $\Omega$ | 0.51 $\Omega$ | 1.05 $\Omega$ |
| $f_0$ | 570 Hz | 570 Hz | 910 Hz | 2200 Hz | 2200 Hz |
| Losses in $C_D$ | 0.26 W | 0.076 W | 0.30 W | 37 W | 0.042 W |

In Example 2 (second-order filter) the total installed capacitance ($C_1 + C_2 + C_D$) becomes huge and the resulting cut-off frequency is low compared to Examples 3 and 4 (see Table 9). In order to achieve comparable capacitances for Example 2, two alternative filters have been calculated. The first alternative has a larger inductance $L_1$, which reduces the total capacitance remarkably. The transfer function stays the same (same Bode plot, see Fig. 15). The selection of $L_1$ allows an optimization of the components in terms of space required, weight, costs, etc., but in general it is much cheaper to store energy in capacitors than in inductors. The second alternative also has a larger inductance $L_1$ and additionally the attenuation factor was relaxed to 0.01 (instead of 0.004). This modification reduces the total capacitance further and the cut-off frequency becomes higher. However, the drawback is a higher output voltage ripple.

Examples 3 and 4 are both fourth-order filters and they have the same transfer function (see the Bode plot in Fig. 15). The only difference is the placement of the damping circuit. Example 4 (damping circuit in second $LC$ stage) is the preferred solution for two reasons: first, the total capacitance is remarkably smaller, which is a space and cost factor. Second, the power dissipation in $R_D$ is three orders of magnitude smaller! Although losses of 37 W might still be acceptable, it requires a larger element, which needs to be cooled sufficiently.

At first glance, the fourth-order filter is more complex, and therefore it is often considered as non-practical. However, the comparison in Table 9 reveals several advantages. High precision power converters need high bandwidth in order to react rapidly to errors. The closed-loop bandwidth is limited by the output filter cut-off frequency (see the Bode plot in Fig. 15). Therefore, higher-order filters, as presented in Example 4, are the preferred choice. If we compare Example 2 with $L_1 = 100\ \mu H$ with Example 4, there is in total 1.6 times more inductance and 3.5 times more capacitance needed to obtain the same attenuation.

Figure 15 shows the comparison of the five examples listed in Table 9. Note, that for Example 2 the two alternatives with $L_1 = 30\ \mu H$ and $L_1 = 100\ \mu H$ have the same transfer function. Also the Bode plots for Examples 3 and 4 are identical.

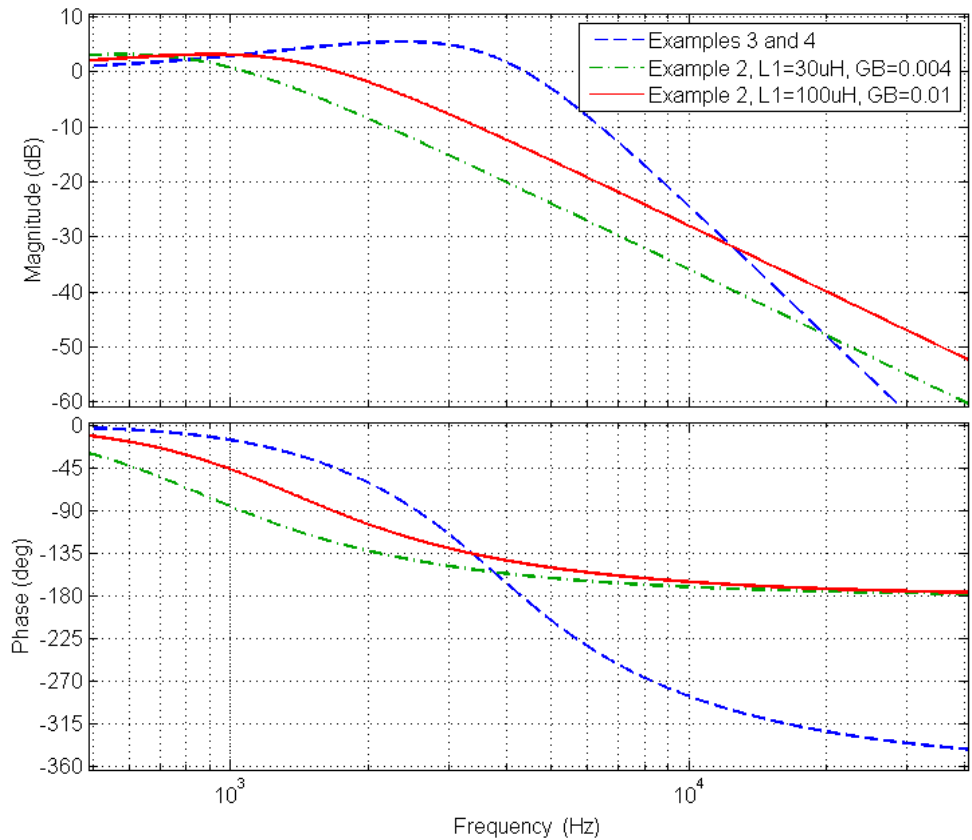

**Fig. 15:** Bode plots for the filter designs according to Examples 2–4 with Bessel optimization

## 5 Practical aspects

### 5.1 Load impedance

The presented calculations do not consider the load impedance. In many cases, where the load impedance is high enough and inductive (magnets), this approach is acceptable. However, if the load impedance is low and/or complex, it has an impact on the filter behaviour. In that case, the load impedance has to be considered in the calculation, or at least the complete circuit has to be analysed.

### 5.2 Parasitic circuit elements

The presented calculations were made considering ideal elements. In practice, this is not the case. As an example, the parasitic resistances and inductances of $C_1$ and $C_2$ in Example 4 should be estimated by considering two alternatives: a 'good' and a 'bad' design. There are two main effects to be considered: the cable that connects $C_1$ and $C_2$ to the circuit, and the ESR (equivalent series resistance) of these two capacitors. In the damping circuit these effects do not have a significant impact.

For the 'bad' design we use a 50 cm long wire with a cross-section of 16 mm$^2$ to connect $C_1$ and a 50 cm long wire with a cross-section of 2.5 mm$^2$ to connect $C_2$. Due to the skin effect, the effective cross-sections of the wires are reduced to 6.3 mm$^2$ for $C_1$ and 2.0 mm$^2$ for $C_2$. These wires add parasitic resistances and inductances of 1.40 mΩ and 0.53 $\mu$H to $C_1$, and 4.2 mΩ and 0.63 $\mu$H to $C_2$.

It is strongly recommended, that $C_1$ and $C_2$ are connected as directly as possible (using shorter connections) to the main bus bars. If we consider connections shorter by a factor of 5 (10 cm) for the 'good' design, the parasitic elements are reduced to 0.27 mΩ and 0.075 µH for $C_1$, and 0.83 mΩ and 0.093 µH for $C_2$.

The ESR values are given in the data sheets. The situation can be improved by selecting good capacitors with a small ESR and by paralleling many small capacitors rather than only a few large ones. The ESR values from a data sheet are given in Table 10 for capacitors suitable for the realization of the output filter according to Example 4 with critical damping (see Section 3.2).

**Table 10:** ESR of metalized film capacitors

| Capacitor | ESR |
|---|---|
| 1.5 µF / 250 V | 6.8 mΩ |
| 10 µF / 250 V | 1.8 mΩ |
| 20 µF / 250 V | 1.9 mΩ |
| 60 µF / 250 V | 1.9 mΩ |

According to Example 4 with optimization method critical damping, $C_1$ should be 124 µF and $C_2$ should be 16 µF. The filter should be realized with capacitors from Table 10.

For the 'bad' design, we use two 60 µF capacitors with a resulting ESR of 0.95 mΩ for $C_1$ and one 20 µF capacitor with an ESR of 1.9 mΩ for $C_2$. Refer also to Table 11.

For the 'good' design, we use 12 10 µF capacitors with a resulting ESR of 0.15 mΩ for $C_1$ and 11 1.5 µF capacitors with a resulting ESR of 0.62 mΩ for $C_2$. Refer also to Table 11.

Figure 16 shows the fourth-order filter as outlined in Section 3.2 expanded with the parasitic elements.

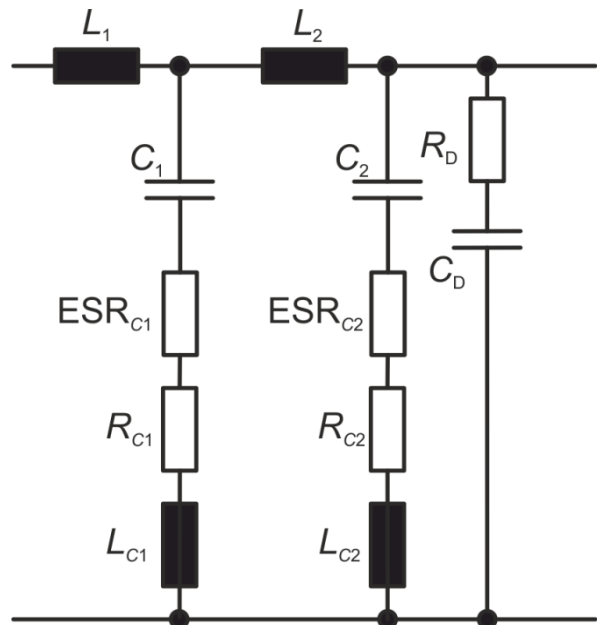

**Fig. 16:** Fourth-order filter with parasitic elements

The corresponding element values are listed in Table 11 for three different designs called 'ideal', 'good', and 'bad'. The main filter elements are rounded to practically realizable values.

**Table 11:** Filter components including parasitic elements

| Parameter | Design 'ideal' | Design 'bad' | Design 'good' |
|---|---|---|---|
| $L_1$ | 30 μH | 30 μH | 30 μH |
| $L_2$ | 17 μH | 15 μH | 15 μH |
| $C_1$ | 124 μF | 120 μF | 120 μF |
| $ESR_{C1}$ | 0 mΩ | 0.95 mΩ | 0.15 mΩ |
| $R_{C1}$ | 0 mΩ | 1.40 mΩ | 0.27 mΩ |
| $L_{C1}$ | 0 μH | 0.53 μH | 0.075 μH |
| $C_2$ | 16 μF | 20 μF | 16.5 μF |
| $ESR_{C2}$ | 0 mΩ | 1.90 mΩ | 0.62 mΩ |
| $R_{C2}$ | 0 mΩ | 4.20 mΩ | 0.83 mΩ |
| $L_{C2}$ | 0 μH | 0.63 μH | 0.093 μH |
| $C_D$ | 382 μF | 360 μF | 360 μF |
| $R_D$ | 0.62 Ω | 0.60 Ω | 0.60 Ω |

Figure 17 shows the Bode plots for the three designs listed in Table 11. For the 'bad' design, there is a steep phase shift right at the switching frequency, and the higher harmonics of the switching frequency are suppressed much less than in the ideal design. For the 'good' design in principle the same occurs, but at higher frequencies and at a much lower amplitude level.

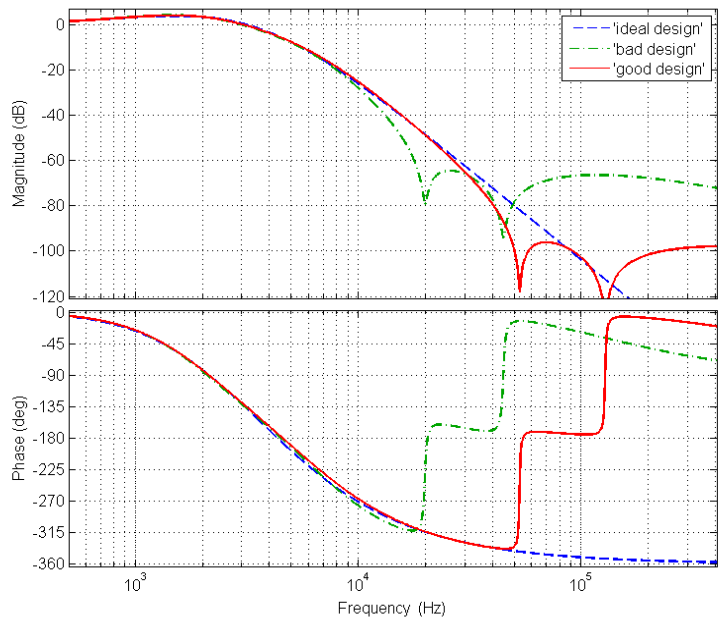

**Fig. 17:** Bode plots for the filter designs with parasitic elements